\documentstyle[aas2pp4]{article}
\input psfig
\newcounter{nref}
\newcommand{\bbib}{%
  \renewcommand{\refname}{\large\bf References}%
  \begin{thebibliography}{99}%
  \setcounter{enumiv}{\thenref}}
\newcommand{\ebib}{%
  \end{thebibliography}%
  \setcounter{nref}{\arabic{enumiv}}}

\begin{document}


\title{Properties of Deflagration Fronts and  Models for Type Ia Supernovae}
\author {I. Dom\'\i nguez $^1$, P.  H\"oflich$^2$}
\affil {$^1$ Dept. de F\'\i sica de Te\'orica y del Cosmos, Universidad de Granada, 
Granada, Spain}
\affil {$^2$ Dept. of Astronomy, University of Texas, Austin, TX 78712 USA}

\begin{abstract}
 Detailed models of the explosion of a white dwarf, which  include self-consistent
calculations of the light curve and spectra, proved a link between observational 
quantities and the underlying explosion model. These calculations assume spherical geometry and
are based on parameterized descriptions of the burning front.
  Recently, first multi-dimensional calculations  
for nuclear burning fronts have been performed. Although a
fully consistent treatment  of the burning fronts is  beyond
the current state of the art, these calculations provided a new 
and better understanding of the physics. 
Several new descriptions for the flame propagation 
have been proposed by Khokhlov et al. and  Niemeyer et al..
Using various description for the propagation of a nuclear deflagration
front,  we have studied the influence on the results of
   previous analyses of Type  Ia Supernovae, namely,
 the nucleosynthesis  and structure  of the expanding envelope.
 
 Our calculations are based on a set of delayed detonation models
with parameters that  give a good account of
the optical and infrared light curves, and of the spectral evolution.
 In this scenario,  the burning front propagates first in a deflagration mode
and, subsequently, turns into a detonation.
 The explosions and light curves are calculated using a one-dimensional 
Lagrangian radiation-hydro code, including  a detailed nuclear network.
 
 We find that the  results of the explosion 
are  rather insensitive to details of the description of the deflagration 
front, even if its speed and the time till the transition to detonation
differ by almost a factor of two. For a given white dwarf (WD) and a fixed transition
density, the total production of elements changes by less than 10 \%, and
the distribution in the velocity space changes by less than 7 \%.
 Qualitatively, this insensitivity of the final outcome of the explosion
on the details of the flame propagation during the (slow) deflagration phase 
can be understood as follows:               
 For plausible variations in the speed of the turbulent deflagration, the duration of this 
 phase is several times longer than the sound crossing time in the initial WD.
 Therefore, the energy produced during the early nuclear burning can be redistributed
over the entire WD causing a slow preexpansion. 
 In this intermediate state the WD is still bound but its binding energy is reduced by
the amount of nuclear energy.
 The expansion ratio depends mainly on the total amount of burning during the deflagration phase.
 Consequently, the conditions are very similar 
 under which nuclear burning takes place during the subsequent
detonation phase.                         
In our example, the density and temperature at the
the burning front changes by less than 3 \% , and the expansion velocity
changes by less than 10 \% . The burning conditions are very
close to previous calculations which used a constant deflagration 
velocity. Based on a comparison with observations, those 
 required  low deflagration speeds ($\approx  $
 2 to 3 \% of the speed of sound). 
 An exception to the similarity
 are  the innermost layers of  $\approx 0.03$ to  $0.05~M_\odot$.
Still, nuclear burning is in nuclear statistical equilibrium (NSE)
 but the rate of electron capture 
is larger for the new descriptions for the flame propagation. Consequently,
the production of very neutron rich isotopes is increased. 
 In our example and close to the center,
$Y_e$ is about 0.44 compared to 0.46 in the model with constant deflagration speed.
This increases the $^{48}Ca$ production by more than a factor of 100 to $3.E-6 M_\odot$. 
 
 Conclusions from previous analyses of light curves and spectra
on the properties of the WD and the explosions will not change and, even
with the new descriptions, the delayed detonation scenario is consistent with
the observations. Namely, the central density, results with respect 
 to  the chemical structure of the progenitor and the transition density 
from deflagration  to detonation  do not change.
 The reason for this similarity is the fact that the total amount of 
burning during the long deflagration phase determines the restructuring
of the WD prior to the detonation. Therefore, we do not expect that the precise, 
microphysical prescription for the speed of a subsonic burning front has a 
significant effect on the outcome.
 
 However, on the current level of uncertainties for the burning front,
 the relation between properties of the burning front and of the
initial white dwarf cannot be obtained from a comparison 
between observation and theoretical predictions by 1-D models.
  Multidimensional calculations  are  needed to get an
inside into the relations between model parameters such as central 
density and properties of the deflagration front, its relation to 
the transition density between deflagration and detonation, and to
make use of information on asphericity that is provided by polarization 
measurements.
 
 These questions are essential to test, estimate and
predict some of the evolutionary effects of SNe~Ia and their use as 
cosmological yard stick.
 
\end{abstract}

\section{Introduction}

 The standard scenario for Type Ia Supernovae  consists of massive carbon-oxygen white
dwarfs (WDs) with a mass close to the Chandrasekhar mass which accrete
through Roche-lobe overflow from an evolved companion star (Whelan \& Iben 1973;
Nomoto \& Sugimoto 1977).
  In these accretion models, the explosion is
triggered by compressional heating.
{}From the theoretical standpoint, the key questions are
how the flame ignites and propagates through the white
dwarf. Several models within this general scenario have been proposed in the past including
detonations, deflagrations and the delayed detonations, which assume that the flame starts as a deflagration and turns
into a detonation later on (Khokhlov 1991,  Yamaoka  et al. 1992,
Woosley \& Weaver 1994). The latter scenario
  and its variation ``pulsating delayed detonation", seems to be the most promising
one, because,  from the general properties and the individual
light curves and spectra, it  can  account for the majority
of SNe~Ia events (e.g. H\"oflich \& Khokhlov 1996, 
and references therein).  
We note that with the discovery of the supersoft X-ray sources,
potential progenitors have been found (e.g. van den Heuvel et al. 1992; Rappaport et
al. 1994).
 
 What we observe as a supernova event is not the explosion itself but the light 
emitted from a rapidly expanding envelope produced by the stellar explosion. As 
the photosphere recedes, deeper layers of the ejecta become visible. A detailed 
analysis of the light curves and spectra gives us the opportunity to determine the
density, velocity and composition structure of the ejecta and provide 
a direct link to       
observations.   A successful application of observational constrains
requires both accurate early light curves (LC) and spectral
observations and detailed theoretical models which are coupled tightly to the 
hydrodynamical calculations (e.g.
Harkness 1991, H\"oflich, Khokhlov \& M\"uller 1991,
 Bravo et al. 1996).
According to previous results, normal bright SNe~Ia 
 can be explained by delayed detonation and  pulsating delayed detonation models (e.g. SN~94D, H\"oflich 1995).
During the deflagration phase, the deflagration velocity is 3 \% of the sound speed. 
In general, a transition from deflagration to detonation is required at densities of about
$2-2.5~10^7~g~cm^{-3}$.
 Central densities of the initial WDs are  $\approx  $
$2.0~10^9 g~cm^{-3}$.
 Despite their success,  the hydrodynamical models are limited by
the parametrized description of  the burning front and the ad hoc adjustment of the 
density at which the deflagration turns into a detonation.
 
Recently,  significant progress has been made toward  a better 
understanding of the propagation of nuclear burning fronts. First 
multidimensional hydrodynamic calculations  of the deflagration fronts have been performed 
(e.g.  Khokhlov 1995,  Niemeyer \& Hillebrandt 1995). 
 A basic, qualitative understanding
of the mechanism which leads to a transition from a deflagration to a detonation phase
may have been achieved
 (Khokhlov, Oran \& Wheeler 1997ab, Niemeyer \& Woosley 1997).
Qualitatively, the results agree between different hydrodynamical numerical simulations but
a full description of the deflagration in the entire white dwarf   and the consistent calculations
of the transition requires high resolution in 3-D that  are beyond the current state of the art.
 Note that the cited references rely on their subgrid models and did not resolve the turbulent
deflagrations on small scales and for appropriate  Reynolds numbers. 
Moreover, the transition from a deflagration to a detonation is still not well understood.
 
 Here, the question is addressed how our results of the explosions vary if we use
descriptions for the deflagration front which use functional relations derived from 3-D calculations.

\subsection { Hydrodynamics}
 
The explosions are calculated using a one-dimensional radiation-hydro
code, including nuclear networks (H\"oflich \& Khokhlov 1996, and
references therein).  This code solves the hydrodynamical equations
explicitly by the piecewise parabolic method (Collela \& Woodward 1984) on
$\approx 1000 $ radial zones. The zones near the burning front are subdivided to
properly track its propagation. The code  
includes the solution of the frequency-averaged radiation transport
implicitly via moment equations, expansion opacities, and a detailed
equation of state. The frequency-averaged variable Eddington factors
and  mean opacities are calculated by solving the frequency-dependent
transport equations.
About one thousand frequencies (in one hundred frequency groups) and
about five hundred depth points are used.  Nuclear burning is 
taken into account using a network of 218 nuclei                  
(see Thielemann, Nomoto \& Hashimoto 1996, H\"oflich, Wheeler \&
Thielemann 1998, and references therein).

\subsection { Description of the Burning Front}

We have considered three cases:
 
Case 1) $v_{burn}= const. ~ v_{sound}$. In our previous investigations, const=0.03 has been
found to give the best fits to observations. This corresponds to the
fractal dimension D=   2 in the description of Woosley \& Weaver (1994)
which suggested $D=2$ to $2.5$.
 
Case 2 \& 3) Here we assumed that $v_{burn}=max(v_{t}, v_{l})~~$ 
where $v_{l}$ and  $v_{t}$ are the  laminar and turbulent velocities, respectively. 
$v_{l}$ is calculated according Khokhlov et al. (1997a).
 
Intrinsically, turbulent combustion is a three- dimensional problem. It is driven
on large scales by the buoyancy of the burning products. The turbulent cascade penetrates
down to very small scales, and makes the rate of deflagration independent of the microphysics.
 Turbulent combustion in a uniform gravitational field and static conditions
 singles out the propagation of the
flame against gravity $g$. Chemical combustion experiments have been performed 
in confined environments, so called
Cumbustion chambers. These experiments can be reproduced by numerical simulations. 
The propagation speed can be described by  
 
$$  v_{t}= C_1~\sqrt{\alpha_{T} ~ g~L_f}; ~~  C_1=0.5; $$
$$\alpha_T ={(\alpha-1)/( \alpha +1}),~\alpha ={\rho^+(r_{burn})/ \rho^-(r_{burn})} ~\eqno{Eq. [1]}$$
 
\noindent
where $\alpha _T$ is the Atwood number, $L_f$ is the characteristic
length scale, and
$\rho^+$ and $\rho^-$ are the densities in front 
and behind the front, respectively.
 However, despite the success in terrestrial experiments, the basic assumptions of both a uniform gravitational field and 
static conditions
are violated in the rapidly expanding envelopes of SNe~Ia. The main effect of expansion  
is the freeze out of the turbulence on scales $L_f$  where the turbulent velocity due to Rayleigh
Taylor instabilities is comparable to the differential expansion velocities on those scales,
i.e. 
       $$v_{t} \approx v_{exp} = L_f/\tau_{ex} \eqno{Eq. [2]}$$
 
 Based on this idea, Khokhlov et al. (1997b) suggested to use the average 
turbulent velocity (eq. 1), use $\alpha $ for uniform, static conditions,
 and to use  the mean expansion time scale 
determined by one dimensional simulations  $\tau_{exp} \approx dt / d~ln ~R_{WD}$. He found
for the propagation speed of the turbulent burning front
 
$$ v_{t}=0.0474*\sqrt{(g~L_f )} \eqno{Eq. [3]}$$
 
 As third case for the description, we followed the recipe of Khokhlov but did some
modifications by taking $\alpha$, $L_f$ and $\tau_{exp}$ directly
from the hydro at the location of the burning front. Freeze-out was assumed 
when the radius of a mass element has doubled after being burned. 
$C_1$ in equation [1] has been varied. Note that a variation in $C_1$
is equivalent to scaling  the relative length scale for the freeze out.
 We varied $C_1$ in the range to cover a  parameter space
which includes both the descriptions suggested by Khokhlov et al.
(1997b) and
Niemeyer \& Woosley (1997).

\begin{figure}[t]
{\psfig{figure=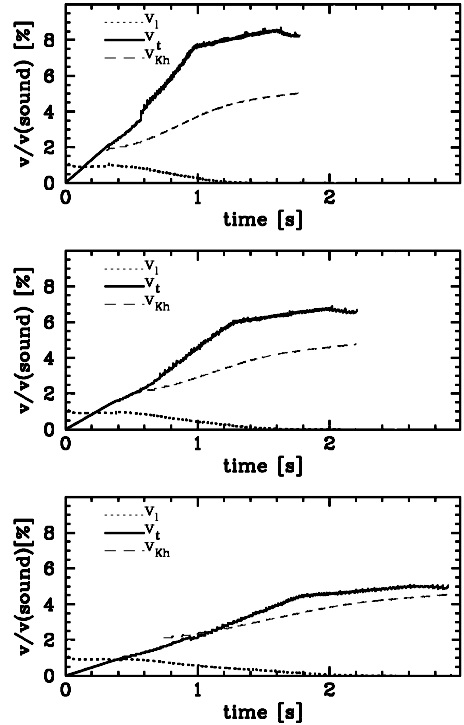,width=6.6cm,rwidth=6.8cm,clip=,angle=360}}
\caption{ 
Laminar   and  turbulent velocities  at the burning front
for models according eq. 1 with $C_1$= 0.25, 0.20 and 0.15
(top to bottom, m2z02y24i3...1).
 For comparison, $v_{Kh}$  gives the velocity of the burning
front according to Khokhlov (eq. 2, m2z02y24i4).}
\end{figure}

\section{Results}                             

The influence of the description of the deflagration front 
 has been studied at the example of a set of delayed detonation model based on 
the same C/O WD with a mass of 1.39 $M_{\odot}$ and a central density
$\rho_c = 2.0~10^9 g~cm^{-3}$. In all cases, a transition density $\rho_{tr}$ of
$2.3~10^7 g~cm^{-3}$ has been assumed.
 The description of the deflagration front has been  varied. 
The deflagration velocity is taken to be 3 \% of the speed of sound 
$v_{sound}$
for {\sl m2z02y24i5},
and  the approximation of Khokhlov is used for {\sl m2z02y24i4}.
 Eq. (1) has been used for  models {\sl m2z02y24i1-3} with $ C_1$=0.15, 0.20 and 0.25.
M2 indicates that the initial mass of the progenitor was 2 $M_\odot$, with solar metallicity (z02),
and 24 \% of Helium.
 
 In figure 1, the velocity of the burning front is shown  as a function of time.
In general, the speed of the burning front is mainly determined by the turbulent
speed but the very early time. As can be expected, the transition density is
reached later in time for smaller $v_{t}$ because the lower energy production
per time and, consequently, the slower pre-expansion.

 The final density, velocity and chemical structures for the most
important elements are given in Figs. 2 
and 3.  Overall, the  structures are very similar because the total
energy release depends on the amount of the released energy and the initial 
structure of the WD. Even the chemical structure or, more precisely, the 
location of transition between different regimes of burning (e.g. from partial to 
total Si-burning) varies between all models by $\leq $ 7 \%  
in the space of the final expansion velocity.
 The total production  of the most abundant elements   
changes by only 4 \% and 3 \% with  mean values of 0.60 and 0.16 $M_\odot$
for $M(^{56}Ni)$ and $M(Si)$, respectively.

\begin{figure}[h]
{ \psfig{figure=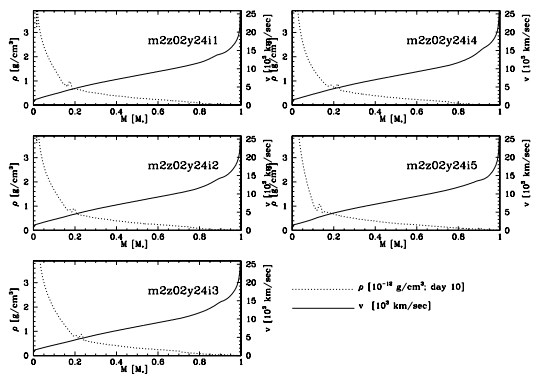,width=7.6cm,rwidth=7.8cm,clip=,angle=270}}
\caption{ 
Final density  
and velocity as a function of mass for different models (see text).}
\end{figure}
\begin{figure}[h]
 {\psfig{figure=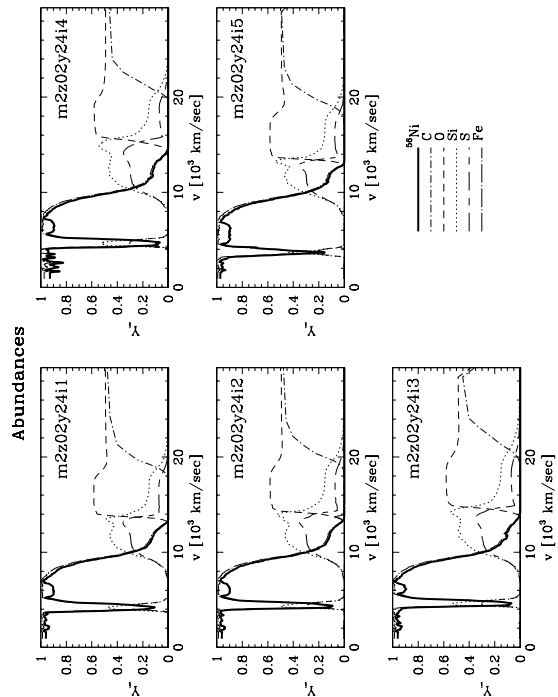,width=7.6cm,rwidth=7.8cm,clip=,angle=360}}
\caption{ 
 {Same as Fig.2 but final chemical composition  as a function 
of mass.}}
\end{figure}
 
 At first,                       
one may expect a rather high sensitivity of the final burning
products on details of the description of the burning front 
(Hillebrandt, Niemeyer \& Woosley
1997) but the tests show a low insensitivity (Fig.2 ... 3).
  This result  can be understood by the very nature of delayed detonation models.
 Qualitatively, the low dependence 
on the details of the flame propagation during the (slow) deflagration phase 
can be understood as follows:               
 The production of intermediate mass
 elements depends on the 
expansion of the outer envelope before the burning front 'arrives'. This pre-expansion occurs 
during the deflagration phase.
 For plausible variations in the speed of the turbulent deflagration, the duration of this 
     phase is several times longer than the sound crossing time in the initial WD.
 Therefore, the energy produced during the early nuclear burning can be redistributed
over the entire WD causing a smooth lifting/preexpansion. 
 In this intermediate state the WD is still bound but its binding energy is reduced by
the amount of nuclear energy.
 The expansion ratio depends mainly on the total amount of burning during the deflagration phase.
 Consequently, the conditions are very similar 
 under which nuclear burning takes place during the subsequent
detonation phase.                         
 In Fig. 4, the burning conditions just behind the front are given for
the two extreme models with $C_1 =$ 0.15 and $C_1 =$ 0.25.
 The durations of the deflagration phase are    
 about 1.7 and 2.9 seconds, respectively.

\begin{figure}[h]
 {\psfig{figure=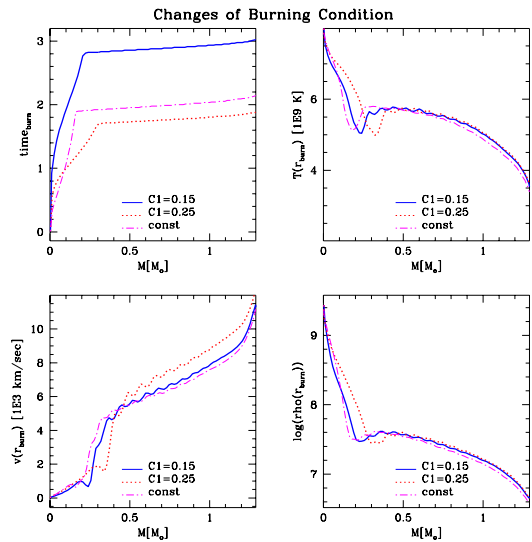,width=7.6cm,rwidth=7.8cm,clip=,angle=360}}
\caption   
{Time at which a mass element is burned, mean expansion velocity, density and  
temperature behind the burning front for  
 deflagration speeds with variable speed according eq. 1 with 
with various $C_1$ (m2z02y24i3/1)
 and with constant deflagration speed (m2z02y24i5).
 We use arithmetic means over 0.05 $M_\odot$.}
\end{figure}
 
 Both models start with laminar deflagrations,
i.e. the conditions under with nuclear burning takes place are
similar. After about 3 \% of a solar mass has been burned, the flame
speed is determined by the turbulent speeds which differ by about 70 \%, 
and the densities behind the burning front      
differ significantly. However, in all models,
the temperatures are sufficient high to burn up to NSE and, mainly,
$^{56}Ni$ is produced. After burning of about 0.25 and 0.32 $M_\odot$ for the
models with low and high turbulent speeds, the transition between the burning
as a deflagration to a detonation is triggered.
 Outside $\approx 0.4 M_\odot $,
 the densities and consequently
the temperatures are very similar.
            Note that the  layers of partial burning are located 
above  $\approx 0.6 M_\odot $.   
The expansion rate, relevant for the time scale for adiabatic cooling, 
differ by 7 \% only. This similarity in the burning conditions explains the
insensitivity in the final chemical profiles.
 The pre-expansion is even more similar than can be expected from this argument
because the hydrodynamical time scales for energy redistribution is $\approx $
1 second.
Thus, the influence of the burning in the deflagration mode
 during the few tenth of a  second before turning to a detonation is reduced with 
respect to the pre-expansion of the outer layers. As soon as the front turns into
a detonation, the remainder of the WD is burned almost 'instantly' or, better,
on nuclear time scales.
 
 For comparison, the properties at the burning front are given for
the description with $v_{burn}/v_{sound} = const.=0.03$
 where the constant
has been tuned to give good agreement between the observed and calculated
light curves and  spectra (e.g H\"oflich \& Khokhlov 1996).
 This description for the burning front is widely used in literature  
and it is consistent with constrains from the nucleosynthesis
(e.g. Brachwitz et al. 1998, Khokhlov 1991, Thielemann et al. 1997, 
 Woosley \& Weaver 1994 with D=2).
 
 In the outer layers, the burning
conditions are virtually identical to $C_{1}=0.15 $ including the expansion 
rate. Slightly less  
material is burned during the deflagration phase because the front is
faster and the initial energy production is larger during the first second
(Fig. 1.).
  
 Note that the amount of burning under very high densities  and, therefore,
the production of neutron rich isotopes in the central region of the WD
 depends sensitively on 
the speed of the front which, in the more realistic descriptions starts 
with laminar speed. In comparison with previous calculations,
this  boosts the production of   
very neutron rich isotopes such as $^{48}Ca$.
 In our example and close to the center,
$Y_e$ is about 0.44 compared to 0.46 in the model with constant deflagration speed.
This increases the $^{48}Ca$ production from $2\times10{-8} M_\odot$  to $3\times 10^{-6} M_\odot$. 
 This increase is the same for all models which start with the laminar deflagration speed.
 For a systematic study of different flame speeds, see the PhD thesis of
Brachwitz (1999). Note that the production of neutron rich isotopes depend on the 
central density and whether the ignition is very close to the center or off-center.
In principle, this opens a new window for detailed analyses of the progenitor and the
ignition process in SNe~Ia.
 
\section {Conclusions} 
 
 The final structure of the expanding envelope
 is rather insensitive to the detailed description of the
burning front during the deflagration phase. It depends mainly on 
global quantities, namely the total amount of burning during the deflagration
phase and the sound speed in the initial white dwarf. Therefore,             
the detailed, spatial structure of the burning front cannot be expected
to change the result of the explosion.                    
 Findings from previous analyses of light curves and spectra
on the properties of the WD and the explosions will not change  and(!) the
new description of the deflagration front is consistent with the observations.
 The central density, conclusions about the chemical structure of the
progenitor and the transition  density from deflagration  to detonation 
do   not change.
 Differentially, evolutionary effects and its consequences for the observations
can be explored by parameterized models.

 The validity of the transition density found in previous analysis
also poses a question on our understanding of the transition from
turbulent deflagration to detonation which is not well understood. Currently, it    is attributed to the
Zeldovich mechanism which is based on the adiabatic mixing of burned and unburned
material where the entropy increase in the unburned fuel triggers the explosion
 (Khokhlov et al. 1997b, Niemeyer et al. 1997). The relative change over the entire
range of parameterizations discussed here
 corresponds to a change in the transition density of
$\approx 10 \%$. This leaves us  with the puzzle why the theoretical estimates
for the transition density are lower by a factor of about 0.7 (Khokhlov et al. 1997b) 
and 0.4 (Niemeyer et al. 1997). For  reasons, we can only speculate. The differences
between the latter values may indicate the size of the uncertainties in our understanding
of this transition process. Other, not yet considered microscopic effects may be involved. Another
possibility may be that we measure two different things. To derive  the model parameters
from the observations, we measure the mean density at the burning front when the transition
occurs whereas the theoretical considerations provide information on the location where
the transition occurs. Maybe the fragmentation of the burning front causes that 
the transition occurs somewhat ahead of the mean front. If this interpretation is correct,
this may indicate that the detonation is started at about 10 to 20 \% (in radius) ahead 
of the mean front. In either case, a clarification needs further investigations
and, certainly,  will provide new inside into the properties of nuclear  burning fronts.

 We note that the nucleosynthesis, optical and IR
 spectra and light curve seem to require that, after an initial phase of slow burning,
the front moves with velocities close to the  speed of sound  to keep up with the expanding 
outer layers and to burn Carbon under low density conditions. The latter is indicated by 
a strong line at 1.05 $\mu m$ which can be attributed to Mg II and indicates expansion velocities
in excess 15,000 ... 16,000 km/sec. In general, 
 unburned Carbon is restricted to  velocities above 20,000 km/sec 
 (Fisher et al. 1997,  H\"oflich 1995, H\"oflich \& Khokhlov 1996, Khokhlov 1991, Nugent et al. 1997, 
Thielemann et al. 1996,  Wheeler et al. 1998). Both the high velocities in Mg and C are consistent
with delayed detonation models but inconsistent with the classical deflagration model W7 (Nomoto et al.
1984).  Extended mixing of the inner, Ni-rich layers can be excluded from the spectra
(H\"oflich, Wheeler \& Thielemann 1998). Mixing of the outer layers may produce Si lines at high
velocity (Nugent et al. 1997) but does not solve the problems with Mg and C. 
Although the evidence favors the delayed detonation scenario in the framework of 1-D models,
this does not necessarily imply that a transition from detonation occurs.
 Alternatively, the flame may propagate as a very fast deflagration wave
 (Hillebrandt, private communication).
 In this context, the transition density may be attributed to the region where the turbulent
flame is not any more driven by Rayleigh-Taylor instabilities.
 
 Finally, we have also to stress the limits of 1-D models with a 
parameterized description of the deflagration front and the need for
multidimensional hydrodynamical calculations and radiation transport.           
 These limits         
are critical to understand     details of the explosion  
 and to use SNe~Ia as distance indicators on cosmological
scales  (e.g. Schmidt et al. 1998, Perlmutter et al. 1999).
 Observations clearly showed a strong relation between the maximum
brightness and the decline relation (Phillips 1993).
 From theoretical models, 
 the amount of radioactive nickel produced during the explosion of a massive white dwarf has
been identified as the basic quantity which provides this relation.
 The relation is independent from details of the model, however, the amount of $^{56}Ni$
actually produced depends on a combination of free parameters in the models such as central density and the 
chemical composition of the WD, and  the propagation of the burning front. If these parameters
are varied independently within the limits indicated by individual fits to observations 
(H\"oflich \& Khokhlov 1996),
we expect a spread around the mean maximum brightness/decline 
 relation of $\approx 0.4^m$ which is consistent with the spread
based on the CTIO data  published by Hamuy et al. (1996). Recent redefinitions of the
statistical methods and new observations suggest a much tighter relation with a spread of
$\approx 0.12^m$ (Schmidt  et. al. 1998).
 This narrow spread cannot be understood in the context of the parameter
range allowed by current analyses.
This tight relation may hint of an underlying coupling of the progenitor, the accretion rates and the 
propagation of the burning front. 
 However, on the current level of
uncertainties, the relation between the free parameters such as transition
density and initial central density cannot be deduced from a comparison 
between observation and theoretical predictions by 1-D models.
  Multidimensional calculations  are  needed to test   
e.g. the  relation between chemical composition or central density of the
white dwarf and the properties of the deflagration front and  its relation to 
the transition density between deflagration and detonation.
 
 In general, the chemical signature of the deflagration phase is 'wiped out'
by the detonation front but a small trace may remain at the interfaces between
complete and incomplete Si burning (Thielemann et al. 1996).
 Although not conclusive, recent polarization 
measurement may already hint toward  the existence of such inhomogeneities
in the chemical structure (Wang, Wheeler \& H\"oflich 1998).                             

 \subsection*{ACKNOWLEDGMENTS}
 
We would like to thank our colleagues A. Khokhlov, J. Niemeyer and
C. Wheeler for helpful discussions and comments.
I. Dom\'\i nguez would like to thank C. Wheeler and his colleagues for the
hospitality during her stay in Austin where this work was started.
 P. H\"oflich  would like to thank  Prof. 
E. Battaner and his colleagues for 
the hospitality during his visit at the University of Granada where 
this work was finished.
 This research was supported in part by  NASA Grant LSTA-98-022,
and the  PB96-1428 of  the Ministery of Education in Spain.
 The calculations were done on a cluster of workstations
financed by the John W. Cox-Fund  of the Department of Astronomy 
at the University of Texas and the High-Performance-Center 
at the University of Texas at Austin.


\end{document}